\documentclass{article}
\usepackage{spconf,amsmath,graphicx,subfigure,hyperref}
\usepackage{tabularx, xcolor}

\title{DEEP AUTOTUNER: A PITCH CORRECTING NETWORK FOR SINGING PERFORMANCES}
\name{Sanna Wager$^{1}$, George Tzanetakis$^{2, 3}$, Cheng-i Wang$^{3}$, Minje Kim$^{1}$}

\address{$^1$ Indiana University, Luddy School of Informatics, Computing, and Engineering, Bloomington, IN, USA\\
$^2$University of Victoria, Department of Computer Science, Victoria, BC, Canada\\
$^3$ Smule, Inc, San Francisco, CA, USA\\}

\begin{document}
\ninept
\maketitle
\begin{abstract}
We introduce a data-driven approach to automatic pitch correction of solo singing performances. The proposed approach predicts note-wise pitch shifts from the relationship between the respective spectrograms of the singing and accompaniment. This approach differs from commercial systems, where vocal track notes are usually shifted to be centered around pitches in a user-defined score, or mapped to the closest pitch among the twelve equal-tempered scale degrees. The proposed system treats pitch as a continuous value rather than relying on a set of discretized notes found in musical scores, thus allowing for improvisation and harmonization in the singing performance. We train our neural network model using a dataset of 4,702 amateur karaoke performances selected for good intonation. Our model is trained on both incorrect intonation, for which it learns a correction, and intentional pitch variation, which it learns to preserve. The proposed deep neural network with gated recurrent units on top of convolutional layers shows promising performance on the real-world score-free singing pitch correction task---autotuning. 
\end{abstract}
\section{Introduction}
\label{sec:intro}
Automatic singing pitch correction is a commonly desired application for digital recordings of singing. However, making a singer's pitch track sound more in tune is not always straightforward. A human listener with a moderate level of musical understanding can often detect the out-of-tune notes and predict the amount and direction of the pitch shift required to bring the note back in tune, all without requiring access to the musical score. However, commercially available pitch correction software depends on a synchronized score for the target pitch \cite{antares:2016}. The lack of knowledge about the target pitch of the sung melody can make a potential automated system suffer in the pitch correction task. We propose a pitch correction program that behaves more like the human ear, basing corrections on information found in the audio, such as the level of perceived musical harmony and context in time. To the best of our knowledge, the proposed method is the first data-driven approach to correcting singing voice pitch based on its harmonic alignment to the accompaniment.

\begin{figure}[t]
    \centering
    \includegraphics[width=\columnwidth]{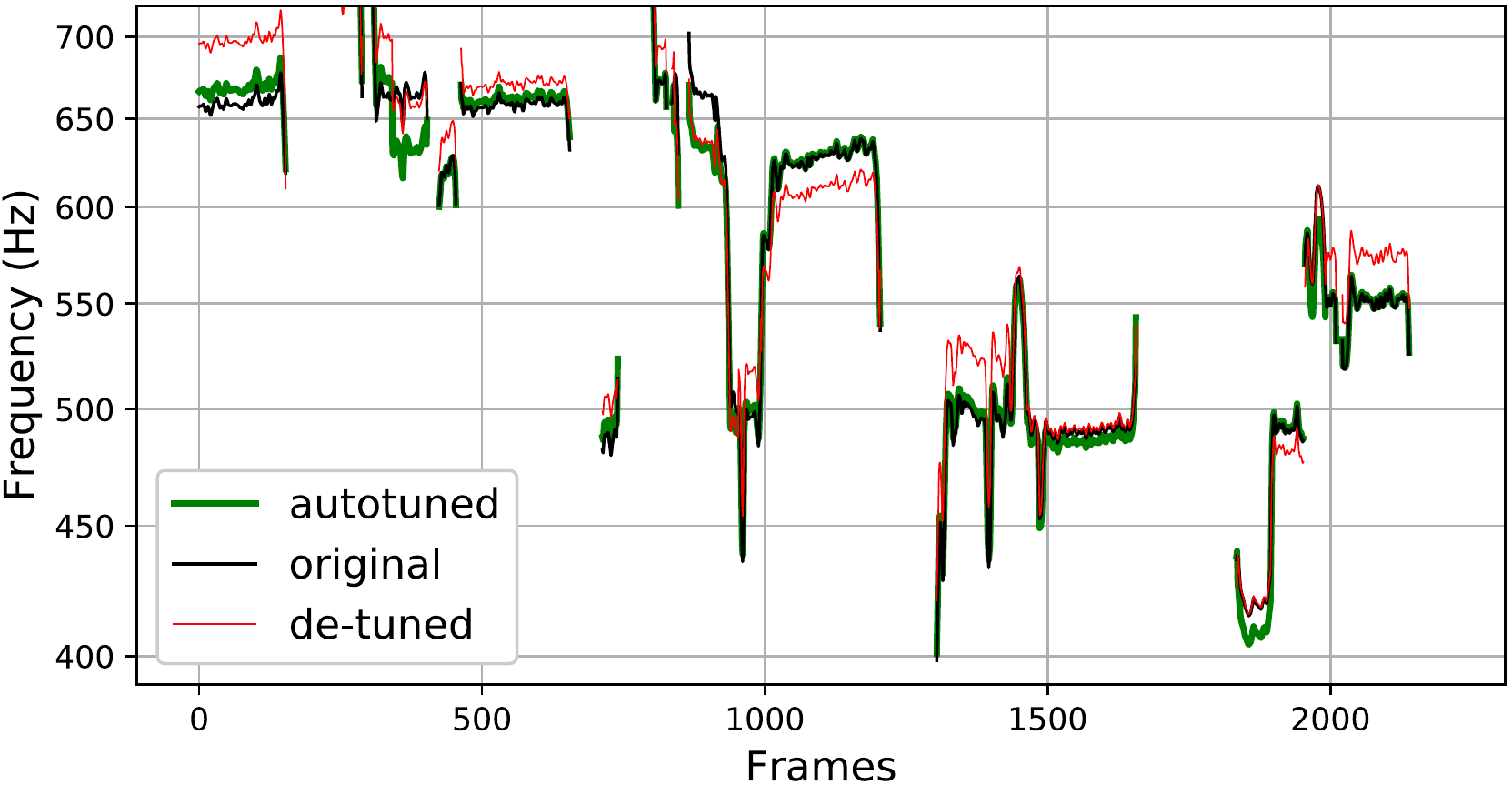}
    \caption{An example of the behavior of the proposed autotuner model.}
    \label{fig:results}
\end{figure}

Fully automatic pitch correction---``autotuning"---is difficult. For example, a score is a sequence of notes of discretized lengths and pre-defined pitches. The simplicity of symbolic representation leaves considerable scope for variation in the singer's interpretation. Hence, although a vocalist follows the general contour of the score, and the result sounds in tune, the singing voice actually varies continuously due to expressive gestures such as pitch bending, vibrato, and any other variations coming from the different genres and personal styles. The proposed data-driven approach tries to respect the nuanced variations of sung pitch, while the system also actively estimates the amount of unintended pitch shift (see Fig. \ref{fig:results}).

\section{Related Work}

The first commercial pitch-correction technique, Antares Auto-Tune \cite{antares:2016}, is also one of the most commonly used. It measures the fundamental frequency of the input monophonic singing recording, then re-synthesizes the pitch-corrected audio signal. In Auto-Tune and in recent work on continuous score-coded pitch correction \cite{salazar2015continuous}, the vocals can either be tuned automatically or manually. In the automatic case, each vocal note is pitch shifted to the nearest note in a user-input set of pitches (scale) or to the pitch in the score if it is known. In the manual case, a recording engineer uses the plugin's interface to move each note to the desired score and precise pitch. With either approach, the default musical scale is the equal-tempered scale, in which each pitch $p$ belongs to the set of MIDI pitches $[0, 1, ..., 127]$ and its frequency in Hertz is defined as $440*2^{\frac{p-69}{12}}$. Some users prefer a finer resolution and include more than twelve pitches per octave, or use intervals of varying sizes between pitches. 

In all cases, the fundamental frequency is discretized to a small set of values, around which every note is shifted to be exactly centered. Hence, the pitch shifts tends to ignore a singer's intentional expressive gestures. To avoid this issue, the user-adjustable ``time-lag" parameter can correct pitch in a gradual way, while introducing a tradeoff between preservation of pitch variation and accuracy. Furthermore, the Auto-Tune system is not easily adaptable to non-Western music with different scales or more fluidly varying pitch. Our proposed model uses a score-free automatic approach, representing pitch as a continuous instead of a discrete parameter.

Recent style-transfer-based work modifies amateur performances to mimic a professional-level performance of the same song. Luo \textit{et al.} proposed to match the pitch contour of the professional-level performance while preserving the spectral envelope of the amateur performance \cite{luo2018singing}. Meanwhile, Yong and Nam proposed to match both the pitch and amplitude envelopes \cite{yong2018singing}. Our model is similar in the sense that it also uses features gathered from high-quality performances \cite{wager2018intonation}. However, the proposed model does not necessitate a ``target" performance of the same song during testing. Instead, it learns from many in-tune singing voice examples and their accompaniments, and then generalizes to unknown songs, while preserving the original singer's style.

Meanwhile, quantitative and qualitative studies on musical intonation show that professional-level singers and instrumentalists often center their frequencies at values that deviate from the equal-tempered scale. This phenomenon is described in \cite{parncutt2018psychocultural}, but dates back to work in \cite{barbour1938just, schoen1926pitch, cameron1907tonal}. Furthermore, soloists often center their singing at a higher frequency than the accompaniment \cite{kantorski1986string, rakowski1985perception}. Devaney {\it et al.} \cite{devaney2011intonation} measure much variety in musical interval sizes both above and below equal-tempered intervals in the case of melodic intervals---where pitches are sequential in time---and polyphonic choral music performed by professional-level singers. Furthermore, frequency and perceived pitch are often slightly different \cite{parncutt2018psychocultural}. The proposed system accommodates this variety of frequencies by letting the fundamental frequency take any value along a continuous scale. 

Gomez {\it et al.} \cite{gomez2018deep} describe recent work on deep learning for singing processing. In Music Information Retrieval (MIR) and speech applications, various combinations of convolutional and recurrent neural networks have been successfully adopted. In \cite{basaranmain}, a convolutional gated recurrent unit (CGRU) estimates the main melody in polyphonic audio signals in the constant-Q transform (CQT) representation, where the gated recurrent unit (GRU) layer \cite{chung2014empirical, ChoK2014arxiv} models temporal structures. The convolutional layer structure is based on \cite{bittner2017deep}, a model for polyphonic pitch transcription on harmonic CQT (HCQT). Since our pitch correction task is sensitive even to a small amount of pitch shift, we choose to use the CQT for its finer log-scale frequency resolution. 



%

\begin{figure*}[t]
\subfigure[]{\includegraphics[height=1.7in]{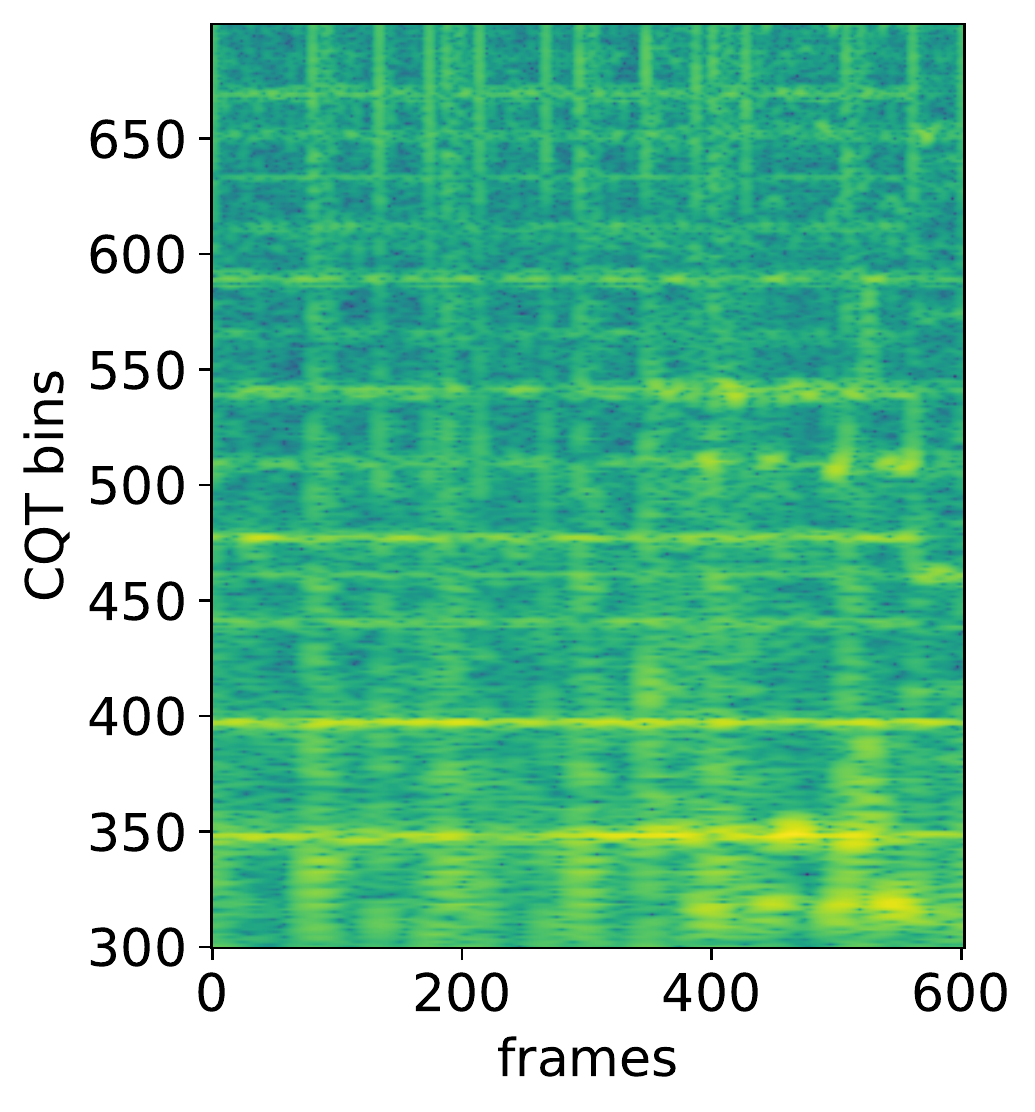}}\vspace{-.03in}
\subfigure[]{\includegraphics[height=1.7in]{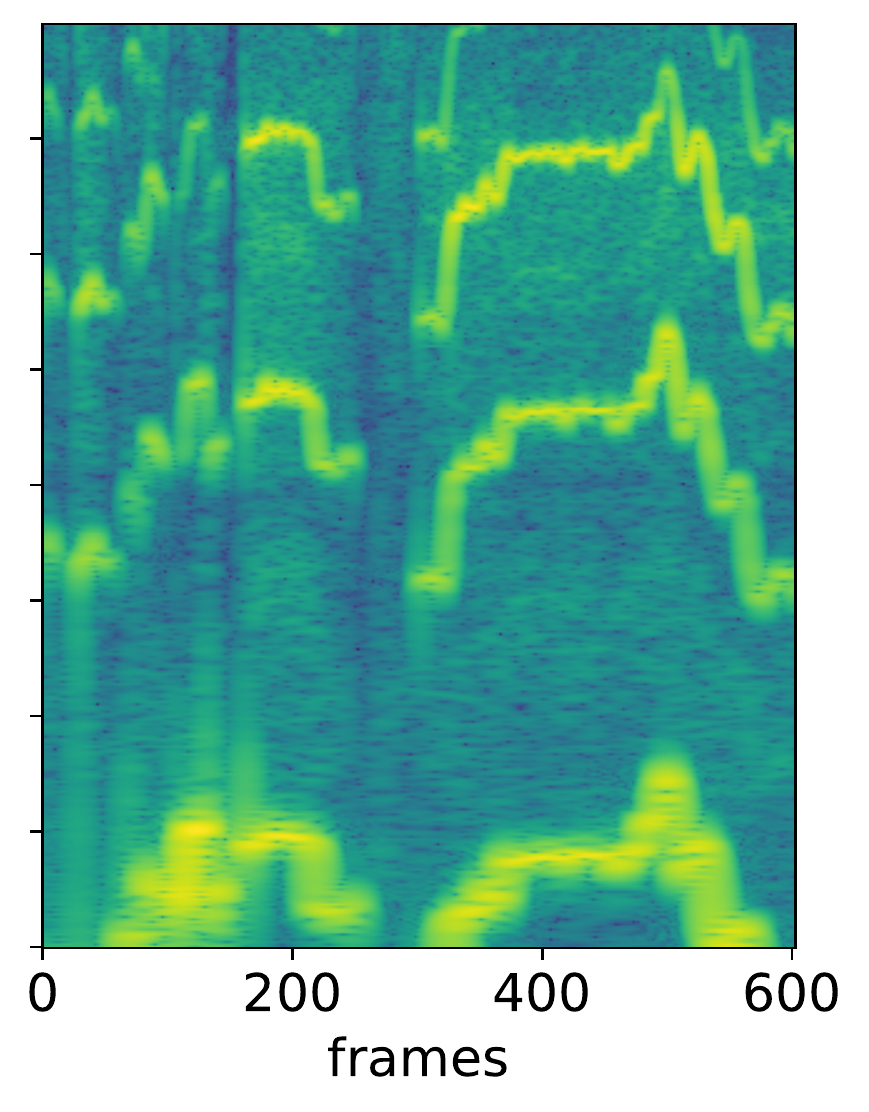}}\vspace{-.03in}
\subfigure[]{\includegraphics[height=1.7in]{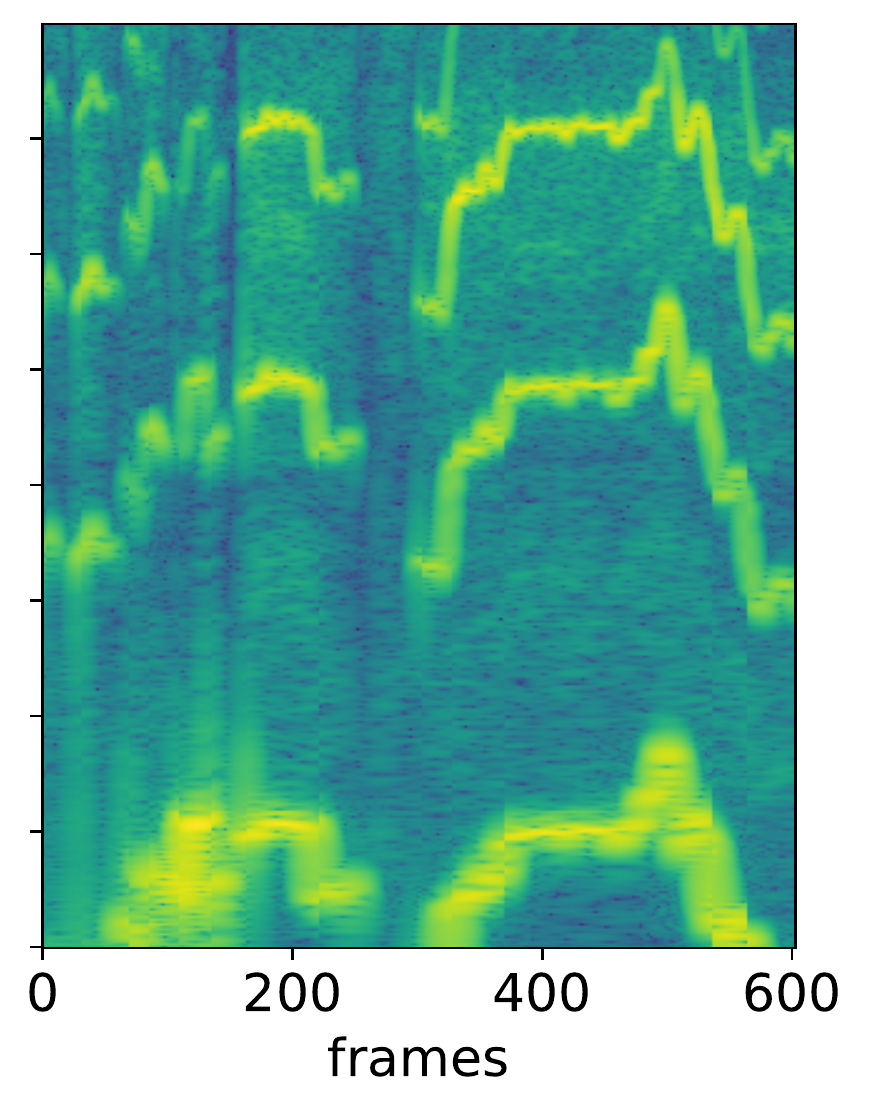}}\vspace{-.03in}
\subfigure[]{\includegraphics[height=1.7in]{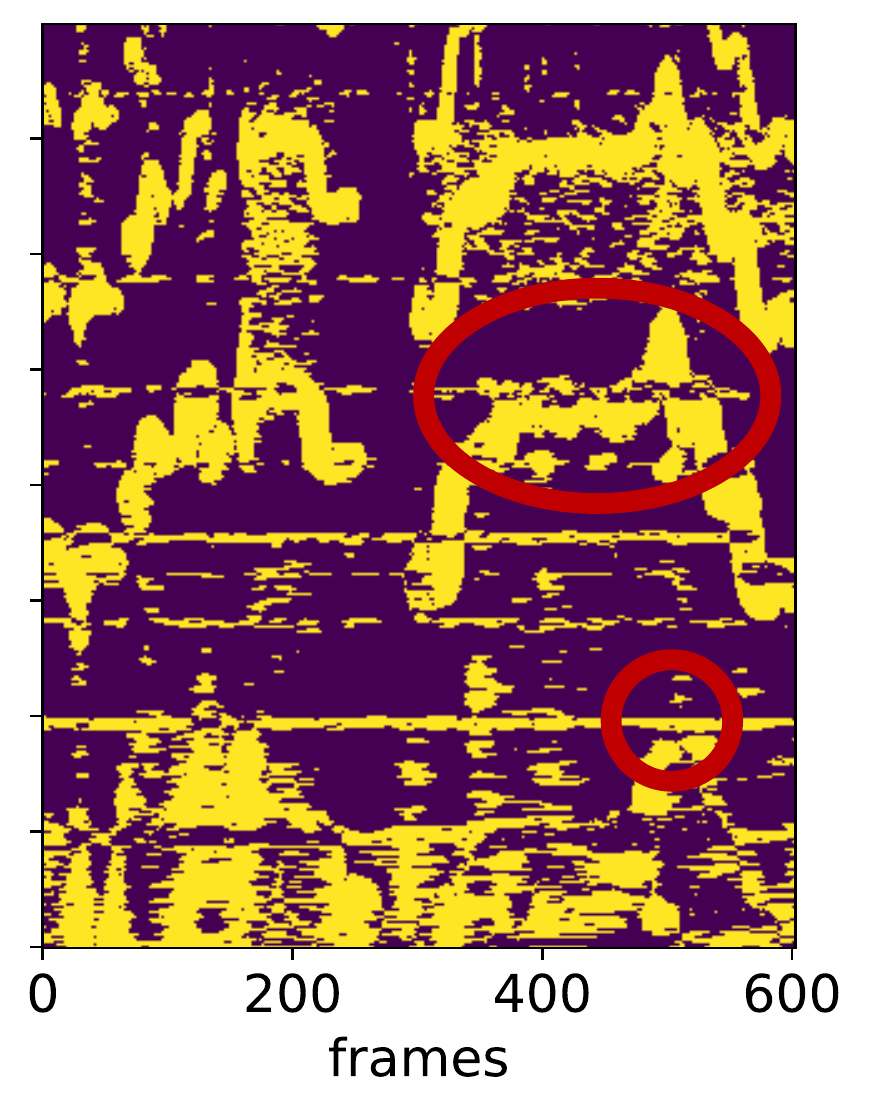}}\vspace{-.03in}
\subfigure[]{\includegraphics[height=1.7in]{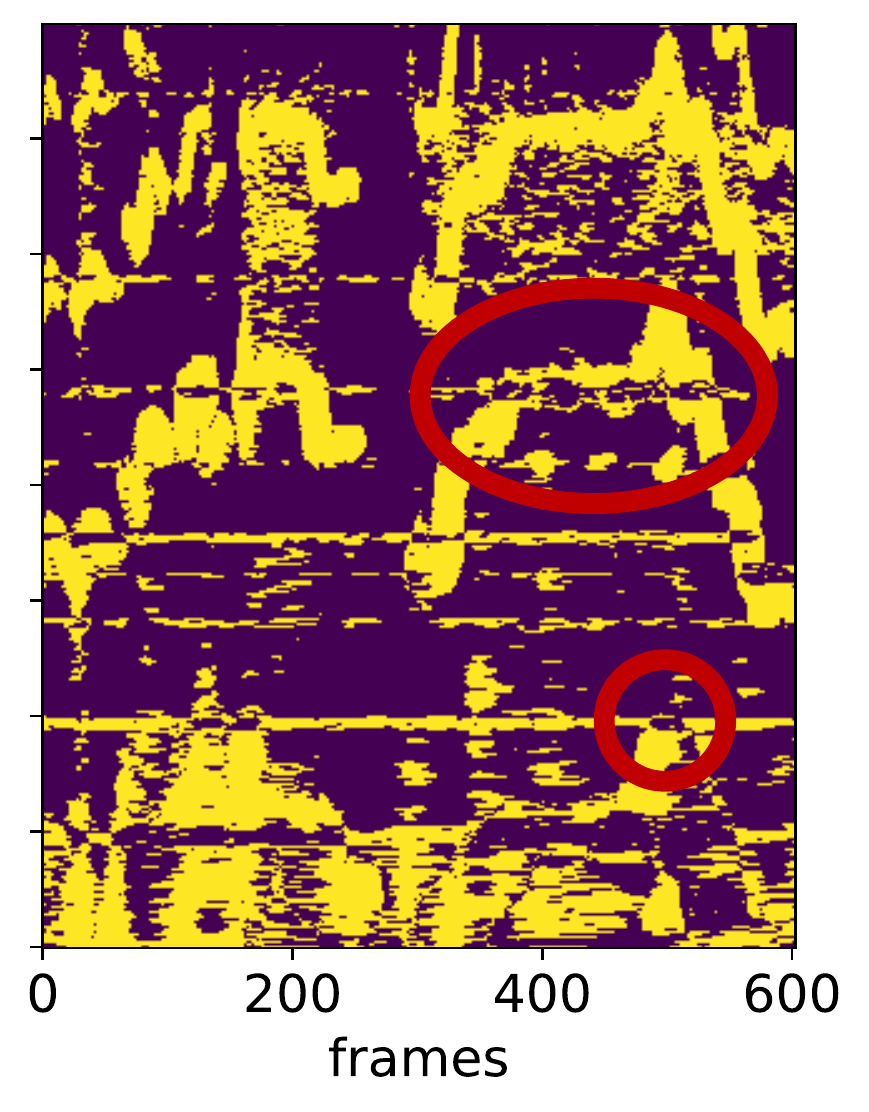}}\vspace{-.03in}
    \caption{
    Constant-Q transform of the vocals and backing tracks using \cite{mcfee2015librosa}. We zoom into bins 300 through 700 for better visibility. (a) shows the CQT of the backing track. The horizontal lines are due to constant pitches. (b) and (c) show the CQT of the vocals before and after the correction, respectively. (d) and (e) show the binarized CQT of the superposed vocals and backing track before and after corrections (see Section \ref{sec:data-format}). The correction shifted the pitch of the vocals up and centered it around the desired harmonics of the backing track (red circles). 
    }
    \label{fig:model-input}
\end{figure*}

\section{The Proposed Pitch Correction System}
\label{sec:proposed}
Our proposed model predicts pitch correction based on the harmonic alignment between the vocals and backing tracks. We assume that the backing track has clearly identifiable pitches---a chord progression---which serves as a reference for the vocals. We make the strong assumption that a singer targets a specific frequency per note, around which all pitch variations are centered. Given this assumption, we correct the pitch of one note by shifting all frames that belong to that note by a constant.

A time-frequency representation is a natural approach for extracting information about the harmonics across a sequence of notes. Fig. \ref{fig:model-input} shows the CQT of vocals and backing track clips before and after applying predicted pitch corrections. It also shows how these shifts affect the harmonic structure of the two tracks combined. Given such data, the proposed model uses convolutional neural network (CNN) layers followed by a GRU to extract sequential patterns of harmonic alignment. It predicts pitch shifts that will align the vocals with the backing track. 

Training data consists of pairs of performances that are identical except for the vocals pitch. Such pairs, while required to train the model, are difficult to come across naturally. Hence, we synthesize them by detuning high-quality singing performances to construct the input signals, and then train our model to predict the shifts that recover the original pitches. The accompaniment track remains fixed.  

\subsection{Dataset}
\label{sec:dataset}
We construct our training dataset by deriving from the ``Intonation" dataset \cite{wager2018intonation}, which we assume to be a collection of in-tune singing voice tracks. The 4702 separate vocal tracks in the dataset are mostly of Western popular music, collected by Smule, Inc, a singing app, for good intonation. While these real-world recordings contain some artifacts, no particular signal processing---e.g. denoising or filtering---has been applied to them. Each recording contains one minute of a performance, starting 30 seconds into the song. Although they are assumed to be in tune, this is not always exactly the case as the users are not necessarily professional singers. Overall, the sung pitch is quite accurate compared with the intended pitch. Hence, we treat this paper as a proof of concept. The model can be trained on professional singing for best results.

Using the metadata indicating the backing track and user index, we split the dataset into 4561 training performances, 49 validation performances, and 64 test performances. The training set contains 709 backing tracks performed by 3468 different users, while the validation set is with 17 tracks sung by 43 users and the test set is with 16 sung by 62. There is no overlap in the backing tracks across the three sets. We allow for overlap in the singer ID between the training and validation sets, but not with the test set. 

We also create another real-world test set using the test backing tracks for a subjective listening test. Outside of Smule, we recorded 8 volunteers singing along with them. Singing experience ranged from beginner to semi-professional. The singers chose what to sing, and selected a total of 7 different arrangements. We recorded a total of 24 performances. Singers familiarized themselves with their chosen songs before the recording session. During the performance, they listened to the backing track through headphones so that it would not interfere with the vocals recording. 

\subsection{The detuning process}
The synthetic pitch deviations used to construct training examples are limited to one semitone (100 cents) in either direction, a larger interval than the standard score-free approach of snapping to the nearest pitch, which limits the shift to 50 cents. In practice, it prevents errors in cases where the required shift is greater than 50, but can lead to degradation of the prediction accuracy on a too badly detuned input. We make the strong assumption is that the detuning process between notes is independent.

The first step in detuning is to find the note boundaries from the original singing performance as our program does not utilize a musical score. We define every transition silence as a note boundary. To this end, we analyze the vocals pitch using the frame-wise probabilistic YIN (pYIN) algorithm \cite{mauch2014pyin}, implemented as a Vamp plugin in \cite{cannam2010sonic}. We then shift every note by a random amount along the continuous logarithmic scale of cents. We generate 7 detuned versions of each song. While our note parsing technique fails to split notes when they are connected, we assume this is not a big problem during training, because the ground truth notes are all assumed to be in-tune. This means that when they are detuned together, the same shift will apply to both. In testing, we used a different technique to compute note boundaries, as described in Section \ref{sec:experiments}.

To detune the training data, we shift the magnitude CQT up or down. This is expected to not produce too noticeable artifacts that the program could learn instead of the pitch relationships. The one issue is formant shifting, but this is not a big concern when only shifting up to $\pm$100 cents. 

\subsection{Data format}
\label{sec:data-format}
We convert the normalized audio signals using the CQT for its translational invariance along the frequency axis. The CQT process covers 5.5 octaves with a resolution of 16 bins per semitone. The lowest frequency is 125 Hz. The top and bottom 16 bins are used as a buffer for pitch shifting, then truncated so that every input has dimension 1024. We use a frame size of 92 ms and a hop size of 11 ms. The vocals and accompaniment CQTs form two of the three input channels to the neural network. For the third channel, to contrast the difference between the first two channels, we binarize the two CQT spectrograms using the mean modulus as a threshold, a technique used in computer vision \cite{sezgin2004survey}. We then take the bitwise disagreement of the two matrices based on the assumption that the in-tune singing voice, better aligned with the accompaniment track, will cancel out more harmonic peaks than the out-of-tune tracks. Fig. \ref{fig:model-input} illustrates the data format. In future work, we plan to explore different input formats, including omitting the binarized channel.

\subsection{Neural network structure}
Our model consists of stacked convolutional layers followed by a GRU layer. The last output of the GRU is fed to a dense layer that predicts a single scalar output, the note-level pitch shift. The convolutional filters pre-process the spectrogram tensor, reducing its dimensionality while also learning abstract features. Next, we keep the last output of the GRU to reduce the representation of a variable-length note to a fixed-length vector.

We use the GRU recurrent structure as a way for the model to analyze the singer's note contour, which can last a second or multiple seconds, while smoothing over unpitched or noisy sections. This is crucial because the algorithm is expected to rely on aligning harmonics, which only occur in pitched sounds. Another advantage of using the GRU is that we can use the hidden state output by one note to initialize the hidden state for the following note. Even when using our simple detuning model that shifts every pitch by an independent amount, we assume that some information from past notes (e.g. from the accompaniment track) is useful.

\begin{table}[t]
  \begin{center}
    \caption{The proposed network architecture.}
    \begin{tabular}{|c||c|c|c|c|}
    \hline
      & Conv1 & Conv2 & Conv3 & Conv4 \\
      \hline
      \#Filters/Units & 128 & 64 & 64 & 64 \\
      Filter size & (5, 5) & (5, 5) & (3, 3) & (3, 3) \\
      Stride & (1, 2) & (1, 2) & (2, 2) & (1, 1) \\
      Padding & (2, 2) & (2, 2) & (1, 1) & (1, 1) \\
      \hline
      & Conv5 & Conv6 & GRU & FC \\
      \hline
      \#Filters/Units & 8 & 1 & 64 & 1 \\
      Filter size & (48, 1) & (1, 1) & & \\
      Stride & (1, 1) & (1, 1) & & \\
      Padding & (24, 1) & (0, 0) & & \\
      \hline
    \end{tabular}
    \vspace{-.1in}
    \label{tab:network}
  \end{center}
\end{table}

Code to both the proposed and baseline results are public\footnote{\href{https://github.com/sannawag/autotuner}{https://github.com/sannawag/autotuner}}. Table \ref{tab:network} displays the structure of the proposed network. Given that the input is a spectrogram, its meaning is different along the time and frequency axes, unlike images. For this reason, instead of max pooling, we use strides of two in the time axis in three of the convolutional layers. In the third layer, we also stride along the frequency axis, but perform this only once to not lose too much frequency information. The fifth convolutional layer has a filter of size 48 in the frequency domain, which captures frequency relationships in a larger range of the CQT, as shown to be successful in \cite{bittner2017deep} and \cite{hsu2017learning}. The error function is the Mean-Squared Error (MSE) between the pitch shift estimate and ground truth over the full sequence of notes in a performance. The MSE corresponds to deviation in cents using the formula $\left|\text{cent error}\right| = 100 * \sqrt{\text{MSE}}$.

\section{Experiments}
\label{sec:experiments}

We use the Adam optimizer \cite{kingma2014adam}, initialized with a learning rate of 0.00005 $5e-5$. We feed one note at a time to the model as a minibatch of seven differently pitch-shifted versions. We apply gradient clipping \cite{pascanu2013difficulty} with a threshold of 100. The convolutional parameters are initialized using He \cite{he2015delving}, and the GRU hidden state of the first note of every song is initialized using a normal distribution with $\mu=0$ and $sd=0.0001$. We measure validation loss every 500 songs and save the model with the best result. 

We apply predicted pitch corrections to test performances. Unlike in training, where we shifted the magnitude CQT and ignored the phase, we use the more computationally expensive TD-PSOLA algorithm \cite{charpentier1986diphone} to detune the signals in the time domain. Furthermore, we compute note boundaries using the pYIN note-level pitch analysis, which splits notes even when there is no silence, but which we consider more prone to errors than our conservative approach in training. In future work, we plan to explore other pitch processing techniques in training, such as \cite{waloschek2018driftin}.

\subsection{Results on the synthesized set}
Our MSE was reduced to 0.049 for training data, 0.062 for validation data, and 0.077 for test data, which corresponds to 22, 25, and 28 cents, respectively. We can conclude that the proposed model works on unseen signals if the detuning behavior is similar to that of the synthetic training signals. In the next section, we test our model on real-world singing signals.

\subsection{Subjective listening test}
\label{sec:subjective-test}

Given that we do not have ground-truth shifts for real-world performances, we use a subjective test to evaluate the proposed model's performance. Listeners compared the original performances to those pitch-shifted using the proposed autotuner. They also compared both versions to the output of a baseline pitch correction system. We designed the baseline to shift each note to the nearest equal-tempered scale degree. Given its design, the program applies shifts of at most 50 cents. We used note-wise pYIN and TD-PSOLA for note parsing and shifting like we did for the proposed model, so the only difference between the outputs is the amount of shift.

We randomly sampled 4 12-second clips from each of the 24 test performances---original and pitch shifted---controlling for vocals presence in 70\% of the frames. We lowered this threshold if four such clips could not be found. We then faded the recording in and out during the first and last seconds. The process resulted in 96 performance clips with 3 versions per clip. The audio samples are publicly available\footnote{\href{https://saige.sice.indiana.edu/research-projects/deep-autotuner/}{https://saige.sice.indiana.edu/research-projects/deep-autotuner/}}.

We generated a blind paired comparison test. Each listener was randomly assigned 10 clips per test but could take the test more than once. For each clip, the listener was randomly assigned 2 out of the 3 versions for comparison. For example, a listener might compare the baseline output of a clip to the original performance clip. They could listen to the samples as much as they wished and restrict the playback to a smaller window. We asked people to select the version they found more accurate, referring to harmonic alignment between the singing voice and the backing track. We also asked them to select the version they found more natural, but received the feedback that users heard no difference, confirming that our TD-PSOLA implementation usually produced clean results. At the beginning of the test, we included links to the original performances of each featured song and two examples to listen to in advance. The test was voluntary and anonymous. Listeners were not aware of the three different cases being compared, only that they were evaluating comparative pitch quality. 10 different subjects with formal musical training provided a total of 138 responses. 

We created a ``quiz'' question to make sure that responses for each quiz were valid. In this question, the original was paired with a version where every note was shifted by a random amount up to 100 cents, and selected to sound noticeably out of tune. All participants answered correctly, which may be due to the fact that they were all musically trained. A musically untrained person who gave feedback on the test did not answer the question correctly.

\subsection{Results on the subjective listening test}
\label{sec:subjective-results}
Globally, in pairs where the subjects compared the proposed program to the original performance, they selected the proposed 33 times and the original 35 times, for a success rate of 49\% with a one-sided binomial test p-value of 0.45. The baseline versus the original produced numbers 24 and 34, or 41\% success rate with p-value 0.12. The proposed model versus the baseline produced 31 and 29, or a success rate of 52\% with p-value 0.45. These global numbers themselves do not suggest an advantage in using either pitch corrector. 

However, in our small sample, we found subjects might prefer proposed autotuned signals when the quality of the original singing is slightly off key, but not too far. We identify such cases by looking at note-level pitch deviation statistics (in cents) between the original performance and the ground-truth MIDI score, generously provided by Smule, Inc. For example, we computed the standard deviation of the cents differences between the original singing and the ground-truth MIDI score. We found that the subjects usually favor an autotuned example if the original was within a particular standard deviation range (between 40 and 60 cents). 19 out of 24, or 79\% of the preferred autotuned examples were in this range, compared to only 16\% (3 out of 19) preferred original examples. While we would need more data for statistically significant results, we expect this behavior due to the imperfectly tuned nature of our crowdsourced training data used as in-tune ground truth---only some noticeable amount of off-pitch can be fixed by the trained model. Meanwhile, the model is exposed only to up to a semitone pitch shift, suggesting that too much variation in the test signal cannot be fixed, either. 

We also compared the proposed method to the baseline. In 18 out of 21 or 86\% of performances where the proposed model was selected, the median of the absolute value of deviations in cents within two semitones was less than 46. However, 8 out of 20, or 40\% of performances where the baseline was selected were in the same range. This second result again that the proposed model might work better when performances are already relatively accurate. See Table \ref{tab:result} for a summary. 

\section{Conclusion}
This experiment is the first iteration of a deep learning model that estimates pitch correction for a monophonic vocal track using the instrumental accompaniment track as reference. Our results on a CNN with a GRU layer indicate that spectral information in the accompaniment and vocal tracks is useful for determining the amount of pitch correction required at a note level. In the future, we plan to move beyond the strong assumptions our initial prototype is based on, and develop a model that predicts (a) corrections without relying on estimated note boundaries (b) the pitch-shifted signal directly rather than the amount of pitch shift (c) corrections more robust to various real-world singing styles and different music genres once more data is available. 


\begin{table}[t]
  \begin{center}
  \vspace{-0.05in}
    \caption{The subjective test results that contrast different distributions of the autotuned, baseline, and original examples. Section \ref{sec:subjective-results} describes the ranges in question.}
    \begin{tabular}{|l||c|c|}
    \hline
      & Within range & Out of range \\
      \hline
      Preferred autotuned examples & 79\% & 21\% \\
      Preferred original examples & 16\% & 84\% \\
      \hline
      Preferred autotuned examples & 87\% & 13\% \\
      Preferred baseline examples & 22\% & 78\% \\
      \hline
    \end{tabular}
    \vspace{-0.2in}
    \label{tab:result}
  \end{center}
\end{table}

\newpage
\bibliographystyle{IEEEbib}
\bibliography{refs}

\end{document}